\documentclass[12pt]{article}
\usepackage{graphicx}
\usepackage{url}
\begin{document}

\def\b{\bigskip}
\def\s{\smallskip} \def\c{\centerline}
\baselineskip=15 pt 
\c{\bf Are classical probabilities instances of quantum amplitudes?}
			\c {Italo Vecchi}
		 \c{ Vicolo del Leoncorno 5 }
		 \c{    I - 44100  Ferrara } 
		\c{    vecchi@isthar.com } \b 
{\bf Abstract:} {\sl A conjecture about the quantum nature of classical probabilites is set forth and discussed.} 
\b

In this note the analogy between classical probabilities and quantum amplitudes is explored . Our theoretical standpoint is based on  ideas presented in a series of previous papers by the author ([1], [2], [3]) , but our argument is  self-contained. Our main  claim is that classical probabilities are instances of quantum amplitudes.  According to such  claim, classical probabilities, refering to "macroscopic" events and "macroscopic" objects, encode information about the  quantum amplitudes of the objects, fudged by the observer's incapacity or unwillingness to track their phases. We also discuss  experimental procedures to verify/disprove that it is indeed so, i.e. to check  wheter our claim is correct.\s

In order to formulate our argument, we must first clarify the meaning  of  some key words. Asking questions and providing answers may be an effective rhethorical device to accomplish this task. \s

{ \bf 1) }   What are probabilities? \s

There are different available interpretations of probability. Here we consider probabilities as a measure of the observer's ignorance (i.e. lack of information) about the state of a system. Such a "subjectivist" interpretation of probability, essentially due to DeFinetti,  is shared by a substantial number of probabilists (see [5] for a survey of  different approaches to probability and a for a good account of the subjectivist interpretation). \s

A key assumption in our argument  is that the evolution of any physical system is accurately described by the laws of quantum mechanics, i.e. by a Schroedinger equation. In particular we assume that the evolution of any isolated system is unitary. Evolution ceases to be unitary if information is extracted from the system. Since information is always relative to an observer's reference frame, this may be reformulated as saying that any system will evolve unitarily as long as no physical exchange breaks the "Heisenberg cut" between the observer  and the system. This point is crucial and so it may be worth elaborating upon.  No perception, i.e. no measurement relative to a system, can take place without a physical interaction between the observer and the system. This means that no perception/measurement can take place without violating the unitarity of the system's evolution. We accept that measurement/perception induces the reduction of the state vector and violates unitarity. \s

It is worth stressing here  the ambiguities of the concept of "environment", which plays a key role in some theories of quantum measurement ([4]).  Information can be defined only with respect to an observer and its reference frame. The very distinction between a system and its environment is actually observer-dependent. As a matter of fact, as implemented by mainstream quantum physicists, the distinction between a system and its environment (incuding the measurement apparatus) coincides with the "Heisenberg cut". The "Heisenberg cut" is relative to the observer and it may be moved at leisure, as long as it leaves the observer perceptory devices (e.g. the observer's retina) on one side and the system on the other. \s

{ \bf 2) } What are mixed states? \s

Confusion about the meaning of  "mixed states" (a.k.a."mixtures") is at the core of some fundamental misconceptions . The mixed-state notation (diagonal matrices etc.) refers to the  observer's lack of information about the system's phases. The mixed state notation encodes information about the possible measurement outcomes on the system relative to a given reference frame ("the pointer basis" of decoherence theory).  In the same way as classical probabilities do not encode events, but only the observer's knowledge/ignorance about the event, the mixed state notation encodes the observer's knowledge about the possible measurement outcomes relative to the system. The distinction between pure and mixed states reflects the observer's incapacity or unwillingness to gather full information about the state of the  system.  \s

{ \bf 3) } What does "macroscopic" mean? \s

In practice "macroscopic" is used as a tacit shorthand for "relative to the observer's perceptions".  In the quantum literature "macroscopic states" are usually those directly associated to the observer's sensory perceptions (i.e. to those measurements where the measurement apparatus coincides with the observer's sensory apparatus). For such perceptions classical physics often provides an accurate description. Furnaces however, whose shine is governed by Planck's radiation law, provide a good example of "macroscopic states" requiring a quantum description ([1]). \s

Once we have somewhat clarified these points, the next step appears both obvious and outlandish.We have just claimed that the classical probabilities relative to measurement outcomes encoded by the mixed-state notation arise from the observer's ignorance about the system's phases. We also know that, as long as the system is not observed, it evolves as a quantum system.  This means that any system, as long as te observer cannot / does not extract information about its state, can be imbedded in a quantum system which evolves unitarily. On this basis it is tempting to make the following conjecture.\s

{ \bf Conjecture.} {\sl Classical probabilities encode the observer's lack of knowledge about a system's quantum phases.} \s

The conjecture does not refer just to so-called "mixed states" of so called "quantum systems". It refers to all physical systems, from Geiger counters to blackjack tables. Essentially we are claiming that all systems are quantum systems and that classical probabilities arise from quantum amplitudes mangled by the observer's inability to determine their phases. According to our conjecture any system whose state is not known by the observer subsists as a superposition. We claim that a coin being thrown in a dark room subsists a superposition of head and tail states as long as the observer does not switch the light on and observes it , inducing state vector reduction. \s

 Heuristically we are induced to make this claim by the analogy with the mixed vs. pure state situation. Along as the system's state is costantly measured by the observer, probabilities are redundant. Indeed, as long as the observer's has full knowledge of the system's state, i.e. as long as his ignorance about the state of the system is zero, the probability of finding the system in a certain state is either zero or one. On the other hand, if the system evolves as a quantum system as long as it is not measured, uncertainty about the possible outcomes of a measurement is encoded by its amplitudes relative to the obsever's reference frame. Since probablities and amplitudes play essentially the same role in classical and quantum systems respectively, it is quite natural to conjecture that probabilities are just amplitudes with unknown phases. The outlandish implication of such a natural guess however is that any kind of uncertainty is a quantum phenomenon. It may be objected that in order for our claim to make sense, the state vector  associated to a system must depend on the knowledge that the observer has acquired about it. Indeed it is so, but, as stated below, with an important proviso. The key difficulty at this point is that if a system is measured by two observers with different degress of knowledge, their measurement outcomes may not coincide. \s
Let consider as an example this sequence of events. \s \begin {enumerate} \item Observers A and B agree on conducting an experiment. \item Observer B throws a coin in a dark room.  \item Observer B switches the light on and observes the state of coin. \item	Observer B switches the light off and exits the room.  \item Without exchanging any information with observer B, observer A enters the room,   switches the light on and observes the state of the coin.  \item Observer A exits the room. \item An information exchange between the observers A  and B concerning the state of the coin ensues. \end{enumerate}\s

One might claim that the state of the coin observed by A and B must coincide and that the state of the coin is  reduced by observer B at stage 3 , even adimitting it was in superposition of tail and head states before being observed.  Such a picture is inaccurate. At stage 3 state of the coin  is reduced with respect to observer B, but not with respect to observer A. 
It appears obvious that observers A and B will agree on the state of the coin that they have seen. Actually without such an agreement scientific exchange, i.e. agreeing on what actually happened, would be impossible. In this setting "reality" is the locus of intesubjective agreement. According to our argument what happened is that at stage 6 both A an B  subsist as superpositions of observers that saw heads or tails. It is only when they exchange information that they agree, i.e that the states of observer A and B collapse respectively with respect to observer B and observer A into states where the coin measurement has yielded the same outcome. We may say that the states of A and B who have seen the same side of the coin are entangled, entanglement being the observer's blueprint for state-vector reduction ([3]). The superposed entangled states of A and B are reduced by B and A so as to preserve intersubjective agreement on experimental outcomes.\s

Some  aspects of our argument are not new, since they have been tentatively set forth in the form or another in the discussion on "Wigner's friend", which is now part of quantum lore. Actually our conjecture can easily be imbedded in Everett's Many Worlds Interpretation of quantum mechanics. (see [6] for a a very readable introduction to MWI)  . What is new here, to our knowledge, is the extension of such an argument to classical probabilities.  Perception-related loss of unitarity has been discussed by other authors ([9]) and intersubjective agreement as a selection rule appears as yet another instance of the "anthropic principle".  Our conjecture inherits all the problematic aspects of the "subjectivist" interpretation of probability, which are mirrored in the ongoing discussion about the observer's role in quantum mechanics.\s 

Since a theory that cannot be tested experimentally is not a scientific theory, the next important question  is whether the validity of our conjecture can be experimentally verified or disproved.
 \s
Essentially the possibility to verify or disprove our conjecture boils down to the possibility to detect superpositions of "macroscopic states" triggered by quantum events. If superpositions of such "macroscopic states" can be detected then our conjecture can be tested quite straightforwardly. All one needs to do is to replace the quantum device used to trigger the macroscopic superpositions with a classical probabilistic device. A concrete instance of such a substitution is replacing the Geiger counter in the original Schroedinger Cat experiment with a "classical probabilistic device", relying, say, on Brownian motion  to induce the desired uncertainty in the system state with respect to an external observer. We stress that when we refer to a "classical probabilistic device" we are resorting to a concept which, if our claim is correct, turns out to be spurious.  If we are right, there is no such thing as a "classical probability": all probabilities are quantum amplitudes. The basic idea here is to prove first that macroscopic superpositions can be detected by showing that they can be distinguished from the states where the system is known (e.g. has been observed) to be in  one of a given set of macroscopic states (e.g. heads or tails)   and then show that macroscopic superpositions behave as states triggered by a classical probabilistic device, where the state of the system is unknown (i.e. has not been observed).  In other words we want first to prove that coins that have been thrown in the air and then observed have physical properties that can be distinguished from those of yet unobserved coins that have been put in a head or tail state by a Geiger counter connected to a relay. Then we want to check that coins which have been "classically"  thrown in the air in a dark room have the same properties as the latter. \b

\begin{tabular} {ll|l}
& & Unobserved Coins thrown  \\
& & by a quantum device \\
& & \hspace{0.5in} $\wedge$ \\
Observed Coins & $<$ {\sl distinguishable} $>$ & \hspace{0.3in}{\sl undistinguishable} \\
& & \hspace{0.5in} $\vee$ \\
 & & Unobserved Coins thrown  \\
& & by a classical device \\
\end{tabular}

\b
In a prevoius paper ([2]) the author proposed an experiment procedure (as yet untested) to detect macroscopic superpositions triggered by a Geiger counter. Replacing the Geiger counter with an appropriate classical probabilistic device  provides the desired testing .\s

It may be noted that superpositions which are arguably macroscopic have been detected in a series of recent experiments. However in such experiments the superpositions are not triggered by a quantum device such as a Geiger counter, which can be straightforwardly replaced by a "classical probabilistic device". It is unclear how such experiments could be adapted to verify or disprove our claim. 
\vfill \eject
		          	     
				\c { \bf References} \b

[1] I.Vecchi, Decoherence and Planck's Radiation Law,  \url{http://xxx.lanl.gov/abs/quant-ph/0002084}\s

[2] I.Vecchi, Interference of macroscopic superpositions, \url{ http://xxx.lanl.gov/abs/quant-ph/0007117} \s

[3] I.Vecchi, Is entanglement observer-dependent?,  \url{ http://xxx.lanl.gov/abs/quant-ph/0106003}  \s

[4] Decoherence and the Appearance of a Classical World in Quantum Theory,  D.Giulini et al. ed. , Spinger, Berlin-Heidelberg-New York 1999. \s

[5] Y.M. Guttmann, The Concept of Probability in Statistical Physics , Cambridge: Cambridge University Press, 1999 \s

[6] M. Tegmark, The Interpretation of Quantum Mechanics: Many Worlds or Many Words?, \url{ http://xxx.lanl.gov/abs/quant-ph/9709032} \s

[7] C.H. van der Wal et al. (2000), Quantum Superposition of Macroscopic Persistent-Current States, Science 290, 773-777  \s

[8] J.R Friedman et al. (2000), Quantum superposition of distinct macroscopic states, Nature 406, 43-46 \s

[9] G.C. Ghirardi (1999), Quantum superpositions and definite perceptions: envisaging new feasible experimental tests, Physics Letters A, 262 (1),  1-14 \s
 
\end{document}